\documentclass[10pt,twocolumn,letterpaper]{article}

 \usepackage{cvpr}              

\usepackage[dvipsnames]{xcolor}

\definecolor{cvprblue}{rgb}{0.21,0.49,0.74}
\usepackage[pagebackref,breaklinks,colorlinks,citecolor=cvprblue]{hyperref}
\usepackage{multirow}
\usepackage{multicol}
\usepackage{enumitem}

\usepackage{booktabs}
\usepackage{siunitx}
\usepackage{arydshln}
\def\paperID{9} 
\def\confName{CVPR}
\def\confYear{2024}

\title{Learning to Classify New Foods Incrementally Via Compressed Exemplars}

\author{
Justin Yang \quad
Zhihao Duan \quad
Jiangpeng He\(^{\dagger}\)  \quad
Fengqing Zhu
\\
Elmore School of Electrical and Computer Engineering, Purdue University, West Lafayette, Indiana, USA \\
{\tt\small \{yang1834, duan90, he416, zhu0\}@purdue.edu}}

\begin{document}
\maketitle
\renewcommand{\thefootnote}{\fnsymbol{footnote}} 
\footnotetext[2]{Corresponding author}

\begin{abstract}
Food image classification systems play a crucial role in health monitoring and diet tracking through image-based dietary assessment techniques. However, existing food recognition systems rely on static datasets characterized by a pre-defined fixed number of food classes. This contrasts drastically with the reality of food consumption, which features constantly changing data. Therefore, food image classification systems should adapt to and manage data that continuously evolves. This is where continual learning plays an important role. A challenge in continual learning is catastrophic forgetting, where ML models tend to discard old knowledge upon learning new information. While memory-replay algorithms have shown promise in mitigating this problem by storing old data as exemplars, they are hampered by the limited capacity of memory buffers, leading to an imbalance between new and previously learned data. To address this, our work explores the use of neural image compression to extend buffer size and enhance data diversity. We introduced the concept of continuously learning a neural compression model to adaptively improve the quality of compressed data and optimize the bitrates per pixel (bpp) to store more exemplars. Our extensive experiments, including evaluations on food-specific datasets including Food-101 and VFN-74, as well as the general dataset ImageNet-100, demonstrate improvements in classification accuracy. This progress is pivotal in advancing more realistic food recognition systems that are capable of adapting to continually evolving data. Moreover, the principles and methodologies we've developed hold promise for broader applications, extending their benefits to other domains of continual machine learning systems.
\end{abstract}



\section{Introduction}
\label{sec:Introduction}
Food image classification is crucial for advancing food pattern tracking, nutritional and health analysis, and dietary monitoring. Despite the high accuracy of current food classification models on static datasets with fixed classes, they falter when faced with the dynamic and evolving nature of food habits, underscored by the frequent emergence of new food categories influenced by changing dietary styles and preferences. This dynamic necessitates that food image classification models are capable of adapting to and learning from continuously changing data. In response to this challenge, our work champions the adoption of continual learning. Continual learning is meticulously designed to incrementally process and learn from data streams, facilitating seamless adaptation. However, a major obstacle encountered within continual learning frameworks is catastrophic forgetting~\cite{Chaudhry_2018_ECCV}, which occurs when models lose previously learned information upon assimilating new data. This phenomenon can significantly degrade the performance of models. Our research focuses on addressing this critical challenge, aiming to enhance the robustness and adaptability of food image classification models in the face of ever-evolving data with the help of compression. This approach holds promising potential for deployment in on-device learning applications, such as food recognition mobile apps, which typically operate within environments under constrained memory and computation resources.

\begin{figure}[h]
  \centering
  \includegraphics[width =\linewidth]{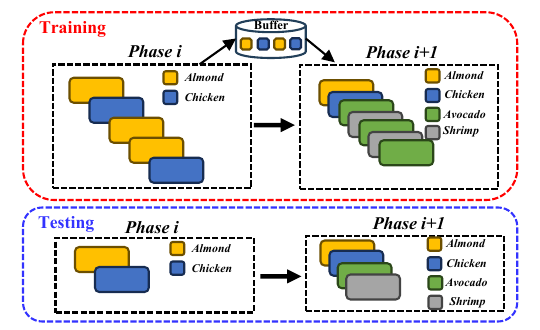}
  \caption{\textbf{Class-Incremental Learning (CIL) for Food Image Classification with Memory Replay}. CIL models progressively learn new food categories presented in a sequential manner.A compact memory buffer retains a subset of previously encountered data, leading to a training dataset that evolves and potentially becomes imbalanced with each incremental training phase. Following each training phase, the performance of the CIL model is based on the classification accuracy across a balanced testing set containing all of the classes it has encountered so far.}
  \label{fig:intro}
\end{figure}

In this work, we mainly focus on class-incremental learning (CIL), which is a more challenging but realistic setting among different continual learning setups~\cite{zhou2023deep}. In CIL, a model is trained on a sequence of tasks, with each task introducing new classes that were not present in previous tasks. During the inference phase, the model is expected to perform classification on all classes seen so far. Memory replay is one of the most effective methods in mitigating the catastrophic forgetting issue in class-incremental learning~\cite{PODNet, iCaRL, mnemonics, Liu2021RMM, lucir, he2024gradient}. It assumes that there is a pre-defined memory budget that allows for storing selected exemplars from previously seen classes. This ensures that the training of incremental learning models incorporates not just new incoming data but also a small set of old samples, commonly referred to as the exemplar set or memory buffer. The objective of this work is to design a continual food recognition system by extending memory replay-based approaches in CIL as shown in Figure~\ref{fig:intro}.

In line with recent advancements in replay based CIL systems, as explored in~\cite{iCaRL}, our framework is structured for preserving a fixed number of previously encountered data within a memory buffer. Under this structure, where the total memory budget is usually limited to a small number of exemplars (such as 20 images per class). The limited set of exemplars may not sufficiently represent the diversity and variability inherent to each class. This lack of representativeness can severely impair the model's ability to generalize well to unseen examples of the class, leading to decreased accuracy. Also, a significant data imbalance issue arises between previously seen classes and new classes. This results in models skewed towards learning knowledge predominantly from new classes, inducing catastrophic forgetting in previously learned classes. Though the intuitive approach to mitigate this imbalance is to increase the memory buffer size, the inherent memory restrictions of CIL present a formidable challenge in expanding this data volume without breaching the established memory limits. The central idea of our method is to leverage image compression techniques, which can play a pivotal role and offer significant advantages. By compressing exemplars, we can increase the number of previously seen class data stored in the memory buffer. This approach results in a more balanced and diverse training set during each incremental learning step, as the compressed exemplars allow for a larger representation of past classes while occupying less memory. Consequently, the model can learn from a more comprehensive set of examples, mitigating the effects of catastrophic forgetting and enhancing its ability to adapt to new classes while retaining knowledge of previous ones. 

However, directly employing compressed images as exemplars introduces the challenge of domain shift in CIL, a complication echoed in findings from recent investigations addressing various vision tasks~\cite{janeiro2023visual, compression_cil}. Notably, compression at low data rates significantly undermines the neural networks' recognition abilities, primarily due to the discordance between the compressed data and its original counterpart. This domain shift, particularly evident when comparing compressed exemplars to the original uncompressed test data within the CIL framework, leads to a noticeable decline in accuracy for previously learned classes at each incremental phase. To mitigate this discrepancy, we involve neural compression~\cite{balle2016end2end, theis2017lossy,balle18hyperprior}, which has shown the ability to obtain less spatial distortion compared with traditional image compression algorithms~\cite{wallace1992jpeg, Skodras2001jpeg2000}, preserving more of the original image's characteristics during the compression process to enhance the data quality in the memory buffer. Worth mentioning, directly finetuning neural compressors in CIL setup introduces its own set of challenges, such as the backward compatibility issues where models may not be able to decode previously compressed bitstream. Therefore, as suggested in~\cite{Duan2024TowardsBC}, we propose fixed-decoder training throughout later incremental stages. In addition, inspired by~\cite{CIM_CIL}, we preserve the foreground semantic information by using class activation map (CAM) to further mitigate the domain shift issue. 

The main contributions of this work include a continual food recognition framework that incorporates a continually trained neural compressor together with CAM-based background compression to enrich data diversity without sacrificing data quality to enhance classification accuracy. We also investigated different compression methods on the background to showcase the importance of compression for CIL settings. The remainder of the paper is organized as follows. In Section~\ref{sec:related}, we review prior works related to food image analysis, incremental learning, and image compression. In Section~\ref{sec:method}, we introduce our proposed method, which includes the memory settings and how we compress exemplars and plug them into class incremental learning framework. In Section~\ref{sec:experiment}, we present the quantitative experimental results with and without compression with a comparison with recent works. Finally, we discuss the conclusion and potential future works in Section~\ref{sec:conclusion}.

\section{Related Work}
\label{sec:related}
\subsection{Food Image Analysis}
Food image analysis, particularly classification, aims to identify and categorize various food items in images accurately. Leveraging cutting-edge deep learning methodologies, these models are trained to recognize different food categories through their distinct visual characteristics, as highlighted in several key studies~\cite{hassannejad2016food,liu2016deepfood, mezgec2019mixed, mao2021visual, mao2021_nutri_hierarchy}. Typically, the process involves assigning a specific food label to each image, under the presumption that the image contains one single category of food item.
Despite these advancements, a notable challenge persists as these models rely on static datasets for model training as pointed out in~\cite{He_2021_ICCVW, raghavan2024online, he2023long}. This constraint significantly curtails the models' ability to adapt and learn from new food images introduced in sequential orders, thus limiting their practicality in dynamic real-life settings where food consumption in daily life might vary from time to time and incorporate new categories.

\subsection{Class-Incremental Learning}
In class-incremental learning (CIL), training data are presented sequentially where the model will encounter new incoming classes at each incremental phase. During testing, the model's performance is evaluated based on the overall accuracy among both previously seen classes and new classes. Consequently, an effective incremental learning model must strike a balance between preserving learned patterns from prior tasks and capturing the characteristics of new tasks. This is also referred to as the stability-plasticity dilemma~\cite{Stability_Plasticity}, where `stability' denotes the preservation of prior knowledge, and `plasticity' signifies the adaptation to new knowledge.
Memory-replay is one of the representative strategies that has demonstrated effectiveness in retaining previously learned knowledge by storing a fixed number of original images~\cite{iCaRL, EEIL, ILIO, mnemonics, POMPONI2020139}, synthetically generated data~\cite{shin2017continual, cGAN_CIL}, or feature prototypes\cite{He_2022_WACV, he2022exemplarfree} as exemplars. 
These stored exemplars are combined with new data at each new step, assisting in the retention of knowledge from past tasks. However, using original data might not always be feasible due to potentially substantial storage requirements, especially when considering the data's dimensions and memory constraints. 

Recent studies have begun to focus on reducing the memory required per image, with the help of compression~\cite{Wang2022MemoryRW, CIM_CIL, compression_cil}. In this work, we introduced a new CIL framework that can integrate with various existing neural compression algorithms in a plug-and-play manner. To address the challenges of backward compatibility~\cite{Duan2024TowardsBC}, we propose to freeze the decoder while continually training the encoder of the compressor. Furthermore, we utilize the class attention map to selectively compress only the background information, thereby preserving the essential foreground features to mitigate domain shift issue~\cite{compression_cil}. 

\subsection{Image Compression}

Image compression methods are broadly classified into lossless and lossy techniques.
Lossless compression algorithms such as PNG~\cite{boutell1997png} compress data by exploring the statistical redundancy of data in a way that the exact original data can be reconstructed from the bitstream.
In contrast, lossy compression methods achieve even lower bit rates by allowing a certain amount of distortion.
This category includes traditional, hand-crafted methods such as JPEG~\cite{wallace1992jpeg}, JPEG 2000~\cite{Skodras2001jpeg2000}, and WebP.

More recently, the development of learning-based lossy image compression has introduced many new methods to the field.
Most existing methods are based on the compressive autoencoder framework~\cite{balle2016end2end, theis2017lossy}.
Notably, the Hyperprior model~\cite{balle18hyperprior}, which introduced a hierarchical entropy model, has significantly boosted the performance of lossy image compression.
Subsequent works~\cite{minnen2018joint, yang2020improving} have further improved this model, demonstrating the potential of learning-based approaches.

\section{Method}
\label{sec:method}
\begin{figure*}[t]
  \centering
  \includegraphics[width = \linewidth]{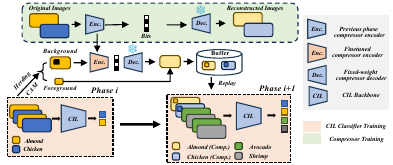}
  \caption{\textbf{Overview of our proposed method.} Our proposed method is divided into two main components:: \textbf{Compressor Training} section and the \textbf{CIL Classifier Training} section. In the compressor training phase, we employ the fix-decoder strategy to train a neural compressor~\cite{minnen2018joint} using only the original data from the current phase. After the CIL classifier training process, we follow standard CIL setups~\cite{iCaRL} using herding to select representative exemplars. These exemplars are then processed through the fine-tuned encoder to generate a compressed exemplar, we also performed CAM-based foreground extraction and then composited them into the the final format in the memory buffer. During the CIL training phase, these compressed exemplar bits are decoded back into images and composite with the foreground image, then incorporated as historical samples within the training workflow, ensuring a continuity of learning that seamlessly integrates past knowledge with new data. Note that light-color images in the figure represent compressed versions of the images (denoted as `Comp.' in the figure).}
  \label{fig:method}
\end{figure*}

In this section, we first introduce the problem setup for class-incremental learning for food image classification, and then we discuss how we train and compress the exemplars using neural compressors to improve memory efficiency in our scenario. Finally, we illustrate how to further mitigate the difference between original and compressed exemplars by combining CAM-based compression with our neural compressor. The overall pipeline for our proposed system is described in Figure \ref{fig:method}.

\subsection{Problem setup for CIL}
\label{subsec:CIL_Intro}
The class-incremental learning (CIL) framework for image classification involves learning a series of $N$ distinct tasks, denoted as $\{T^0, T^1, \ldots, T^{N-1}\}$, where each task contains unique, non-overlapping classes. Each task $T^c$, for $c < N$, consists of $M$ training data $\{\left(\mathbf{x}^c_j, y^c_j\right)\}^M_{j=1}$, with $\mathbf{x}^c_j$ representing the image data and $y^c_j$ representing the label for the $c^{\text{th}}$ task. By incorporating compression into the CIL setup, each compressed image requires less memory and can therefore result in a more diverse and rich exemplar set.

To support retaining previously learned knowledge while learning new information, an exemplar set that contains only past encountered classes is introduced. Denoted as $\mathcal{E}=\left\{\left(\mathbf{x}^p_j, y^p_j\right)\right\}_{j=1}^{L}$, where $p < c$ and $L$ indicates the memory constraint that limits the maximum number of images the buffer can store. This exemplar set, selected via herding~\cite{herding} after the completion of the most recent task's training, is merged with the data for the current task ($\mathcal{E} \cup T^c$), enabling the model to learn new tasks while preserving information from prior ones.

The management of the exemplar set is essential for maintaining the efficiency and effectiveness of CIL systems, ensuring their ability to continuously learn and adapt to new tasks over time without significant degradation of previously acquired knowledge.

In our setup, we employ compressed versions of images rather than original ones as exemplars. Accordingly, we define our exemplar set as $\hat{\mathcal{E}}=\left\{\left(\mathbf{\hat{x}}^p_j, y^p_j\right)\right\}_{j=1}^{L^{'}}$, with $\mathbf{\hat{x}}^p_j$ represents the compressed images. Compressed data enables increased storage capacity in memory buffer, denoted as $L^{'}$, surpassing $L$, the maximum capacity for original images. This enhancement stems from the reduced data rate, typically measured in bits per pixel (bpp), calculated by dividing the total memory requirement by the pixel count. The decreased memory needs of compressed exemplars result in a lower bpp and hence enable a larger amount of compressed exemplars as studied in~\cite{compression_cil}.

\subsection{Continual Neural Compressor Training}
The neural compressor training process is a critical initial step in our proposed method, underpinning the seamless integration of continual learning with efficient data compression. At the core of this process lies the utilization of a fixed-decoder strategy, which is meticulously applied to train a neural compressor. This compressor is trained on the original data available from the current phase, ensuring that the compressor is finely attuned to the specific characteristics and nuances of the most recent dataset. While our framework accommodates any advanced neural compressor backbones, we demonstrate our method using the mean scale hyperprior model \cite{balle18hyperprior}, a classic model serving as the foundation for many existing compression architectures. The essence of employing a neural compressor within our framework is to achieve two objectives: firstly, to substantially reduce the storage requirements for exemplar data in the memory buffer, and secondly, to maintain or even enhance the quality of the data representation. The compressor achieves this by encoding the data into a compact bitstream format, which captures the essential information of the images while discarding redundancies.

The learning objective is to minimize the rate-distortion (R-D) loss:
\begin{equation}
    \label{eq:cloc_background_rd_loss}
    \min \, \mathbb{E}_{X \sim p_{\text{c}}} \left[ - \log_2 p_Z(Z) + \lambda \cdot d(X, \hat{X}) \right].
\end{equation}
Here, \(X \sim p_{\text{c}}\) denote data samples in a single incremental step. For this neural compressor, a neural network encoder, denoted as $E$, maps \(X\) to a latent variable \(Z \triangleq E(X)\), while a neural network decoder, denoted as \(\mathcal{D}\), maps \(Z\) back to a reconstruction \(\hat{X} \triangleq \mathcal{D}(Z)\). A learned probability distribution \(p_Z\) is employed to model the distribution of \(Z\).
\(d\) represents a distortion metric, which we use mean squared error in our experiments, and \(\lambda\) is the Lagrange multiplier that trades off between rate and distortion, which we set as $\lambda = 16384$ following the setting in~\cite{balle18hyperprior}. The minimization is over the network parameters of $E$, \(\mathcal{D}\), and \(p_Z\) in the initial incremental step, and only $E$ in the following steps, as we keep \(\mathcal{D}\) and \(p_Z\) fixed throughout later incremental steps to ensure backward compatibility.

After the training process of the compressor. We then compress the selected exemplars, identified using herding~\cite{herding}, into a bitstream and store them in the buffer for future use. This compressed format is not only space-efficient but also designed to be readily decodable, ensuring that the stored exemplars can be accurately reconstructed into images for subsequent use in the latter CIL training phase.

During the continual learning classifier training, the decoder decodes the bitstream back into images. These decompressed exemplars are then reintegrated into the training pipeline as historical samples. This strategic reutilization of compressed data is what allows our method to effectively bridge the gap between learning new information and retaining previously acquired knowledge, all the while optimizing the use of limited memory resources. Through this innovative approach, our neural compressor training process establishes an efficient and effective way to incorporate neural compressor training into continual learning to adapt the ever-evolving data.
%
\label{subsec:NC_Training}


\subsection{CAM-Based Image Compression}
To further reduce the distortion between our compressed exemplars and the original counterpart. We also performed CAM-based compression on the exemplars. Class Activation Map (CAM)~\cite{CAM} is a powerful technique in computer vision to identify which regions in an image are relevant to a particular class in a classification task, especially in Convolutional Neural Networks (CNNs) in our case. To minimize additional required memory for storing CAM information, we turn CAM into bounding box information by first generate a binary mask from a CAM and then convert it into a bounding box, follow these steps:

\begin{enumerate}
    \item \textbf{Generate CAM}: First, forward an image through a CNN until the last convolutional layer to obtain feature maps. For a specific class, the CAM is calculated by weighting the feature maps with the weights of the fully connected layer corresponding to that class. Mathematically, the CAM for class \(c\) at spatial location \((x, y)\) is given by:
    \[ \text{CAM}_c(x, y) = \sum_{k} w_{k,c} \cdot F_k(x, y) \]
    where \(w_{k,c}\) are the weights of the fully connected layer for class \(c\), and \(F_k(x, y)\) are the activation values of feature map \(k\) at location \((x, y)\).

    \item \textbf{Generate Binary Mask}: Threshold the CAM to create a binary mask. Here we set the threshold \textit{T} as 0.6 in our experiments to ensure that areas with high activation are set to 1 (foreground) and the rest to 0 (background)
    \[ M(x, y) = \begin{cases} 
    1 & \text{if } \text{CAM}_c(x, y) > \textit{T} \\
    0 & \text{otherwise}
    \end{cases}
    \]

    \item \textbf{Convert into Bounding Box}: Finally, to convert the binary mask into a bounding box, find the smallest and largest \(x\) and \(y\) coordinates of pixels in the mask that are set to 1. These coordinates define the corners of the bounding box that encapsulates the region of interest. Specifically, the bounding box \(B\) can be defined as:
    $B = (\min(x_{\text{fg}}), \min(y_{\text{fg}}), \max(x_{\text{fg}}), \max(y_{\text{fg}}))$
    where \(x_{\text{fg}}\) and \(y_{\text{fg}}\) are the sets of \(x\) and \(y\) coordinates of the foreground pixels, respectively.
\end{enumerate}

\noindent Let \(I\) denote the original image, and \(\hat{I}\) represent the compressed version of the original image. The final generated image \(I_{\text{final}}\) is obtained by merging the foreground of \(I\) with the background of \(\hat{I}\), guided by the binary mask \(M\) obtained from the CAM. The final image \(I_{\text{final}}\) at \((x,y)\) is given by:
\[ I_{\text{final}}(x, y) = M(x, y) \cdot I(x, y) + (1 - M(x, y)) \cdot \hat{I}(x, y), \] where \(M(x, y) \cdot I_{\text{orig}}(x, y)\) selects the foreground pixels from the original image, and \((1 - M(x, y)) \cdot \hat{I}(x, y)\) selects the background pixels from the compressed image.
\section{Experiments Results}
\label{sec:experiment}

\subsection{Datasets}
In adherence to the protocol of class incremental learning outlined in~\cite{iCaRL}, three public datasets are used for evaluation, including two real-world food datasets Food-101~\cite{Food101} and VFN-74~\cite{vfn74}, and one general image classification dataset ImageNet~\cite{ImageNet} to showcase that our plug-in method not only works for food image analysis but also works for general image classification tasks. \textbf{Food-101} is a food dataset that contains 101 food categories with 750 training images and 250 testing images per class. \textbf{VFN-74} is a class-imbalanced real-world food dataset that contains a total of 15,000 images across 74 most frequently consumed food classes in U.S.A, and we randomly select 50 images as testing data across all classes. \textbf{ImageNet-100} is a subset of the large-scale dataset ImageNet-1000~\cite{ImageNet}, which contains around 1.2 million training images and 50k validation images, with varying image sizes. It is important to note that ImageNet-100 is selected from the first 100 classes after a random shuffle. Consistent with the protocol in~\cite{iCaRL}, all classes are shuffled using the random seed of 1993.

\subsection{CIL Protocols}
\label{sec:CILProto}
Two different protocols were used in our experiments: learning from scratch (LFS) and learning from half (LFH), following recent CIL works~\cite{Foster, der}. The main difference between LFS and LFH is the number of classes in the first incremental step, which we term as base class size. The step size represents the number of new classes after the initial step. For LFS, the first step classes are equal or similar to step size. For Food-101, we tested on two LFS settings: (Base 11, Step 10) and (Base 21, Step 20). For VFN-74, we tested on (Base 14, Step 5) and (Base 14, Step 10). For ImageNet-100, we tested on (Base 10, Step 10) and (Base20, Step 20). For LFH, the first step contains roughly half of the total classes, which we use (Base 51, Step 5) and (Base 51, Step 10) for Food-101, (Base 34, Step 5) and (Base 34, Step 10) for VFN-74, and (Base 50, Step 5) and (Base 50, Step 10) for ImageNet-100. Another difference between LFS and LFH is the memory setting following the setup in~\cite{Foster, der}. In LFS, we apply the fixed memory setting buffer size of 2,000 original images for Food-101 and ImageNet-100, and 370 original images for VFN-74. For LFH, we apply the growing memory setting to store 20 original images for each seen class for Food-101 and ImageNet-100, and 5 original image per class for VFN-74 due to the scope of the data. For compressed exemplars, the memory calculation setup follows~\cite{compression_cil} to use bitrate be pixel (bpp) to calculate equivalent memory.

\subsection{Implementation Details}
\textbf{Compared Methods: }For comparison, we aim to select several representative class incremental learning methods that use exemplar sets for training, including iCaRL~\cite{iCaRL}, WA~\cite{wa2020}, PODNet~\cite{PODNet}, DER~\cite{der} and FOSTER~\cite{Foster}. In addition, we plugged CIM~\cite{CIM_CIL} in DER and FOSTER and compared it with the results by integrating our exemplar compression framework.   

\noindent\textbf{Training Details: }The methods outlined are implemented using PyTorch~\cite{pytorch}. To ensure a fair comparison, each method utilizes an identical backbone for feature extraction: ResNet18~\cite{resnet}, complemented by a straightforward fully-connected layer acting as the classifier, in line with the standard class-incremental learning (CIL) framework detailed in~\cite{iCaRL}. The training hyperparameters were harmonized across all approaches, employing the Stochastic Gradient Descent (SGD) optimizer with a starting learning rate of 0.1 and a momentum of 0.9. The training duration was established at 200 epochs for the initial phase and 170 epochs for each following phase, with the learning rate scheduled to decrease by a factor of 0.1 after 80 and 120 epochs. A batch size of 64 was maintained throughout, and to ensure reproducibility, a fixed random seed (1993) was applied across all experiments. Furthermore, AutoAugmentation~\cite{Cubuk2018AutoAugmentLA} was systematically applied to enhance the dataset in all experimental setups, adhering to the protocol suggested in~\cite{Foster}. The mean scale hyperprior model~\cite{balle18hyperprior} used in our experiments is trained from scratch with 200 epochs during the first phase, and we finetuned the compressor with 50 epochs using our fix-decoder strategy in the following incremental phases. Note that only original data at each step is used to fine-tune the compressor, and we use Adam optimizer~\cite{Adam} with a fixed learning rate of 2e-5.

\noindent\textbf{Evaluation Metrics: }In the testing phase, the model is evaluated by its accuracy over all previously encountered classes, without the provision of a task index. We present the top-1 average accuracy (Avg) across all learning stages, providing a comprehensive view of the model's performance throughout the entire continual learning process. Furthermore, the top-1 accuracy at the final stage (Last) is also provided, offering insight into the model's effectiveness upon the completion of the continual learning cycle for all observed classes. 

\begin{table*}[h]
    \centering
    \begin{tabular}{
        l
        S[table-format=2.2]
        S[table-format=2.2]
        S[table-format=2.2]
        S[table-format=2.2]
        S[table-format=2.2]
        S[table-format=2.2]
        S[table-format=2.2]
        S[table-format=2.2]
    }
    \toprule
     & \multicolumn{8}{c}{Food-101} \\
    \cmidrule(lr){2-9}
     & \multicolumn{4}{c}{Learn from Scratch} & \multicolumn{4}{c}{Learn from Half} \\
    \cmidrule(lr){2-5} \cmidrule(lr){6-9}
     & \multicolumn{2}{c}{Step Size: 10} & \multicolumn{2}{c}{Step Size: 20} & \multicolumn{2}{c}{Step Size: 5} & \multicolumn{2}{c}{Step Size: 10} \\
    \cmidrule(lr){2-3} \cmidrule(lr){4-5} \cmidrule(lr){6-7} \cmidrule(lr){8-9}
    {Methods} & {Avg} & {Last} & {Avg} & {Last} & {Avg} & {Last} & {Avg} & {Last} \\
    \midrule
    iCaRL~\cite{iCaRL} & 67.29 & 52.43 & 72.66 & 61.98 & 56.04 & 45.56 & 63.32 & 51.98 \\
    WA~\cite{wa2020} & 71.11 & 55.56 & 77.15 & 67.88 & 56.81 & 45.78 & 67.9 & 57.42 \\
    PODNet~\cite{PODNet} & 67.62 & 51.37 & 75.30 & 64.09 & 69.21 & 58.41 & 74.12 & 63.65 \\
    \hdashline
    DER~\cite{der} & 77.83 & 70.82 & 79.36 & 73.43 & 76.61 & 71.34 & 77.68 & 73.35 \\
    DER \textit{w/} CIM & 77.38 & 71.43 & 79.46 & \bfseries 75.04 & 76.77 & \bfseries 72.45 & 78.35 & 74.00 \\
    DER \textit{w/} ours & \bfseries 77.87 & \bfseries 72.11 & 79.46 & 73.35 & \bfseries 76.97 & 71.97 & 72.41 & 65.22 \\
    \hdashline
    FOSTER~\cite{Foster} & 72.60 & 64.07 & 76.92 & 70.58 & 73.00 & 63.58 & 78.76 & 72.38 \\
    FOSTER \textit{w/} CIM & 74.48 & 66.24 & 78.21 & 72.14 & 74.33 & 65.28 & 79.52 & 73.96 \\
    FOSTER \textit{w/} ours & 74.78 & 66.85 & \bfseries 79.50 & 74.16 & 75.36 & 67.73 & \bfseries 79.56 & \bfseries 74.25 \\
    \bottomrule
    \end{tabular}
\vspace{-0.2cm}
\caption{Accuracy (\%) of various Class Incremental Learning (CIL) methods on the Food-101 dataset, contrasting approaches that learn from scratch with those that utilize partial pre-existing knowledge. The table highlights the effectiveness of different methods, including those with our compression technique, in improving model performance under different step size scenarios. Best results are bold.}
\label{tab:Acc_Tab_Food101}
\end{table*}

\begin{table*}[h]
    \centering
    \begin{tabular}{
        l
        S[table-format=2.2]
        S[table-format=2.2]
        S[table-format=2.2]
        S[table-format=2.2]
        S[table-format=2.2]
        S[table-format=2.2]
        S[table-format=2.2]
        S[table-format=2.2]
    }
    \toprule
     & \multicolumn{8}{c}{ImageNet-100} \\
    \cmidrule(lr){2-9}
     & \multicolumn{4}{c}{Learn from Scratch} & \multicolumn{4}{c}{Learn from Half} \\
    \cmidrule(lr){2-5} \cmidrule(lr){6-9}
     & \multicolumn{2}{c}{Step Size: 10} & \multicolumn{2}{c}{Step Size: 20} & \multicolumn{2}{c}{Step Size: 5} & \multicolumn{2}{c}{Step Size: 10} \\
    \cmidrule(lr){2-3} \cmidrule(lr){4-5} \cmidrule(lr){6-7} \cmidrule(lr){8-9}
   {Methods}  & {Avg} & {Last} & {Avg} & {Last} & {Avg} & {Last} & {Avg} & {Last} \\
    \midrule
    iCaRL~\cite{iCaRL} & 66.26 & 51.96 & 71.4 & 60.42 & 52.07 & 50.96 & 59.33 & 56.58 \\
    WA~\cite{wa2020} & 66.65 & 52.94 & 72.2 & 61.72 & 60.26 & 50.92 & 60.8 & 57.84 \\
    PODNet~\cite{PODNet} & 71.03 & 52.39 & 76.5 & 62.74 & 73.29 & 61.94 & 75.86 & 65.7 \\
    \hdashline
    FOSTER~\cite{Foster} & 76.60 & 66.38 & 79.41 & 70.34 & 76.55 & 68.3 & 77.67 & 71.18 \\
    FOSTER \textit{w/} CIM & 77.27 & 67.64 & 80.04 & 71.32 & 76.07 & 68.32 & 78.06 & 71.68 \\
    FOSTER \textit{w/} ours & \bfseries 78.73 & \bfseries 68.22 & \bfseries 80.34 & \bfseries 71.76 & \bfseries 77.01 & \bfseries 69.62 & \bfseries 78.09 & \bfseries 72.56 \\
    \bottomrule
    \end{tabular}
\vspace{-0.2cm}
\caption{Accuracy (\%) of various Class Incremental Learning (CIL) methods on the ImageNet-100 dataset, comparing approaches that learn from scratch with those that leverage partial pre-existing knowledge. Best results are bolded, highlighting the effectiveness of the proposed method when integrating our compression technique.}
\label{tab:Acc_Tab_ImageNet}
\end{table*}


\subsection{Results and Analysis}
\vspace{-0.1cm}

\begin{figure}[t]
    \centering
        \includegraphics[width=\linewidth]{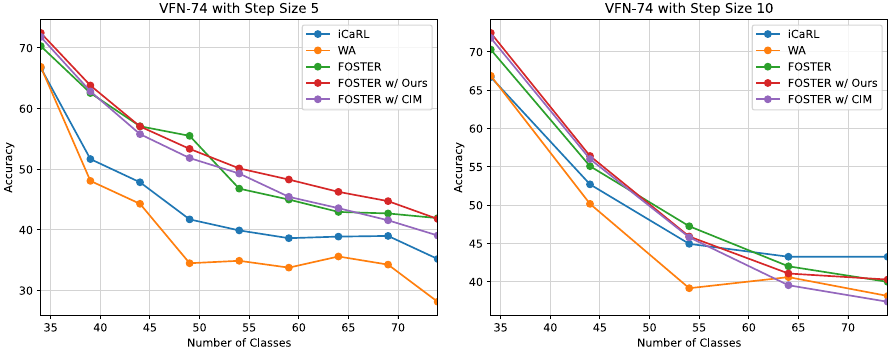}
        \label{vfnfig}
    \vspace{-0.5cm}
    \caption{Results on VFN-74 dataset under LFH setting with exemplar set of 5 images per class using various CIL methods. The incremental step size $M \in \{5, 10\}$. We also included results with plug-in methods CIM~\cite{CIM_CIL} and our method.}
    \label{fig:vfn_graph}
\end{figure}

The impact of integrating our compressed exemplars into two existing benchmark approaches, FOSTER and DER, was explored across three datasets (Food-101, ImageNet-100, and VFN-74) and within two CIL protocols (LFS and LFH), as outlined in Tables~\ref{tab:Acc_Tab_Food101},~\ref{tab:Acc_Tab_ImageNet}, and Figure~\ref{fig:vfn_graph}, respectively. 

Our method consistently enhances classification accuracy across large-scale datasets (Food-101 and ImageNet-100), particularly under smaller step sizes. This improvement is attributed to the compressor's ability to learn a more generalized encoder, effectively reducing both rate (memory usage) and distortion when more current step data is available at each incremental learning phase. However, the fixed batch size and training epoch for the compressor training phase result in fewer training iterations for VFN-74 compared to Food-101 and ImageNet-100. Consequently, VFN-74 experiences less gain in classification performance due to less effective compression. This observation aligns with our ablation study in Section~\ref{sec:ablation}, underscoring the importance of background information in classification. 

The benefits of implementing compression become more pronounced when classifier is trained on sufficient data to be generalizable to different compression artifacts. For Food-101 and ImageNet-100, for each incremental step with step size of 10, it contains 7,500 and 13,000 images respectively, while for VFN-74, the amount of data is around 2,000. Also, the inherent data imbalance issue harms the performance for CIL methods, even when compression is involved. When the data imbalance issue is minimal, such as for Food-101 and ImageNet-100, some CIL methods already exhibit good performance, and the benefits of compression may not be immediately apparent. In such cases, compression techniques may initially sacrifice accuracy in the early phases of CIL. However, the value of compression becomes evident in later CIL phases, particularly when memory constraints are tight, and the need for efficient storage and retrieval of exemplars becomes more critical. By effectively managing the trade-off between memory usage and classification accuracy, our compression method demonstrates its utility in enhancing the performance of CIL systems under resource-limited scenarios.

\begin{table}[h]
    \centering
    \begin{tabular}{lcccc}
    \toprule
    \multirow{2}{*}{Ablation Method} & \multicolumn{3}{c}{ImageNet-100 (LFS)} \\
    \cmidrule{2-4}
     & Avg & Last & \# Exemplars \\
    \midrule
    \textit{w/o} Exemplar comp & 76.55 & 68.3 & 20 \\
    \textit{w/} Blank background & 71.09 & 63.76 & 59.8 \\
    \textit{w/} Background removal & 71.84 & 61.41 & 64.4 \\
    \textit{w/} CIM comp~\cite{CIM_CIL} & 77.97 & 67.74 & 33.7 \\
    \textit{w/} Jpeg comp & 78.01 & 67.36 & 42.5 \\
    \textit{w/} Ours & 78.73 & 68.22 & 45.1 \\
    \textit{w/} Ours + Pretrain & \textbf{79.18} & \textbf{69.2} & 45.1 \\
    \bottomrule
    \end{tabular}
    \caption{Ablation methods to compress exemplars with CAM-based compression with FOSTER~\cite{Foster} backbone. \textit{``\# Exemplar"} here means the average number of compressed exemplars per class with equivalent memory size.}
    \label{tab:Compression_Acc_Tab_comparison}
\end{table}

\begin{figure*}[t]
    \centering
    \includegraphics[width=\linewidth]{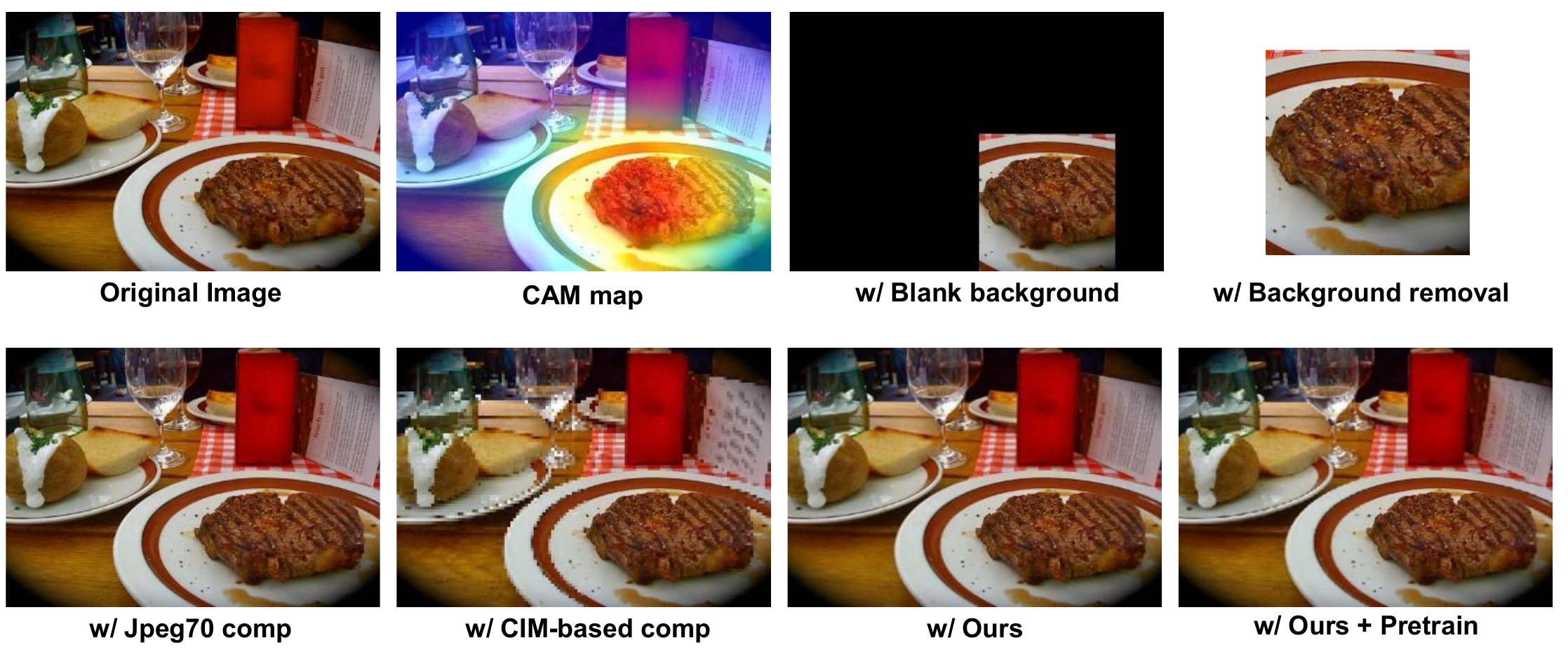}
\caption{Visualization of different compression methods applied to a food image from the `steak' class.}
     \label{fig:steak_visualize}
\end{figure*}
\vspace{-0.1cm}
\subsection{Ablation Study on Background Compression}
\label{sec:ablation}
In this section, we delve into ablation studies to evaluate the efficacy of our class activation map (CAM)-based background compression strategy. The motivation behind employing CAM for background compression is to minimize distortion between the compressed exemplars and their original images. Given the hypothesis that the background might not be essential, and focusing solely on the foreground—where the convolutional neural network predominantly concentrates—might allow for storing more exemplars without distorting the original image's essence, we explored various approaches to background compression, including its complete removal and substitution with a uniform blank. These experiments are meticulously detailed in Table~\ref{tab:Compression_Acc_Tab_comparison}, where we assess the impact on classification accuracy by comparing last-step and average top-1 accuracies and noting the total number of compressed exemplars achievable under the memory constraints equivalent to 20 original images per class.

Our findings reveal that eliminating or blanking out the background adversely affects the neural network model's classification capabilities, despite increasing the size of the compressed exemplar pool. We also examined four distinct compression methodologies: area-based compression as described in CIM~\cite{CIM_CIL}, JPEG compression~\cite{Wang2022MemoryRW}, our CAM-based technique, and a pre-trained neural compressor incorporating our CAM-based background strategy. The visualization of an example image is shown in FIgure~\ref{fig:steak_visualize}. Our method outperforms both CIM-based and JPEG compression in accuracy, albeit slightly trailing behind the pre-trained compressor model. This discrepancy might happen because of the distribution shift when fine-tuning the neural compressor with different category data throughout each incremental phase, and also because of the differences in the volume of training data and the duration of the training period. In neural compression, models typically benefit from extensive training on large datasets, such as 90k images from the Vimeo dataset~\cite{xue2019video}, which is far beyond the scope of our experimental setup (10,000 in ImageNet-100, 8,000 in Food-101 and 3,000 in VFN-74). Such comprehensive training endows models with a generalized capability for encoding/decoding a broad spectrum of images, thereby enhancing rate-distortion performance. While leveraging external data for training is not viable within our experimental constraints to maintain a fair comparison with uncompressed methods, deploying a pre-trained neural compressor in real-world scenarios could be advantageous. It offers the benefit of obviating the need for compressor retraining at each phase and sidesteps the distribution shift issues associated with fine-tuning neural compressors.

\vspace{-0.1cm}
\section{Conclusion and Future Work}
\label{sec:conclusion}
\vspace{-0.1cm}
In real-world ML systems, particularly in food classification, adaptability to evolving data is crucial for incorporating new food categories while avoiding catastrophic forgetting. Our research integrates neural compression algorithms with class-incremental learning (CIL) to enhance data efficiency. We introduced plug-in method that incorporates neural compressors with fixed-decoder training for CIL, specifically addressing food classification challenges. By incorporating a portion of the original data through CAM-based compression, we mitigate domain shifts, enabling the development of efficient, lifelong ML systems. These advances promise more accurate and storage-efficient food classification, crucial for health monitoring and dietary tracking technologies. Future work could explore spatially adaptive compression and classifier-guided compression strategies to further refine CIL performance.

\newpage
{
    \small
    \bibliographystyle{ieeenat_fullname}
    \bibliography{references}
}


\end{document}